\documentstyle[11pt,paspconf,psfig]{article}

\begin{document}

\title{The U.S. ISO\altaffilmark{1} Key Project on Quasars} 

\author{Eric Hooper\altaffilmark{2}, Belinda Wilkes\altaffilmark{2},
Kim McLeod\altaffilmark{3}, Jonathan McDowell\altaffilmark{2}, Martin
Elvis\altaffilmark{2}, Matthew Malkan\altaffilmark{4}, Carol
Lonsdale\altaffilmark{5}, Chris Impey\altaffilmark{6}}

\altaffiltext{1}{ISO is an ESA project with instruments
funded by ESA Member States (especially the PI countries:  France,
Germany, the Netherlands, and the United Kingdom) and with the
participation of ISAS and NASA.}

\altaffiltext{2}{Harvard-Smithsonian Center for Astrophysics, 60
Garden St., Cambridge, MA 02138}
\altaffiltext{3}{Department of Astronomy, Wellesley College, 106
Central St., Wellesley, MA 02181}
\altaffiltext{4}{Astronomy Department, UCLA, Los Angeles, CA 90095}
\altaffiltext{5}{IPAC, Caltech MS 100-22, 770 S. Wilson Ave., Pasadena, CA  }
\altaffiltext{6}{Steward Observatory, University of Arizona, Tucson,
AZ 85721}


\section{Introduction}

A substantial fraction of the bolometric luminosity of many quasars
emerges in the infrared (Elvis et al. 1994), from synchrotron radiation
and dust.  Which of these emission mechanisms is dominant depends on
quasar type and is an open question in many cases.  The non-thermal
component is likely connected with radio and higher frequency
synchrotron radiation, providing information about the relativistic
plasma and magnetic fields associated with quasars.  Much of the dust
emission is due to heating by higher energy photons from the active
nucleus, and is therefore important for understanding the overall
energy balance.  Moreover, this thermal component provides an
orientation-independent parameter for examining unification
hypotheses.  Fundamental phenomenological questions about quasar
infrared spectral energy distributions (SEDs) include:  the range of
SEDs within each quasar type; differences between one type and
another; the evolution of the SEDs; and correlations with fluxes at
other wavebands, host galaxy properties, and orientation indicators.  

The IRAS satellite provided an enormous boon for infrared studies of
quasars, detecting a large number of sources for the first time in the
mid-infrared.  Its major limitations were a lack of coverage of the
far infrared, $\lambda > 100 \mu$m; shallow limiting flux, especially
in the sky survey from which most quasar data were obtained; and
relatively low spatial resolution.  The recently completed ISO mission
(Kessler et al. 1996) was an important step in the evolution of
infrared astronomy, providing extensive spectroscopic capabilities, a
large number of photometric passbands from 3 to 200 $\mu$m, higher
spatial resolution, and the opportunity to achieve deeper flux limits.
Two major ISO observing programs have obtained broad-band photometry
for large samples of quasars: a European Core program which focused on
low-redshift, predominantly radio-loud quasars; and a US Key Project
to examine quasars spanning a wide range of redshifts and SEDs, e.g.,
X-ray and IR-loud, plus those with unusual continuum shapes.

\section{Observations with ISO}

\begin{figure}
\psfig{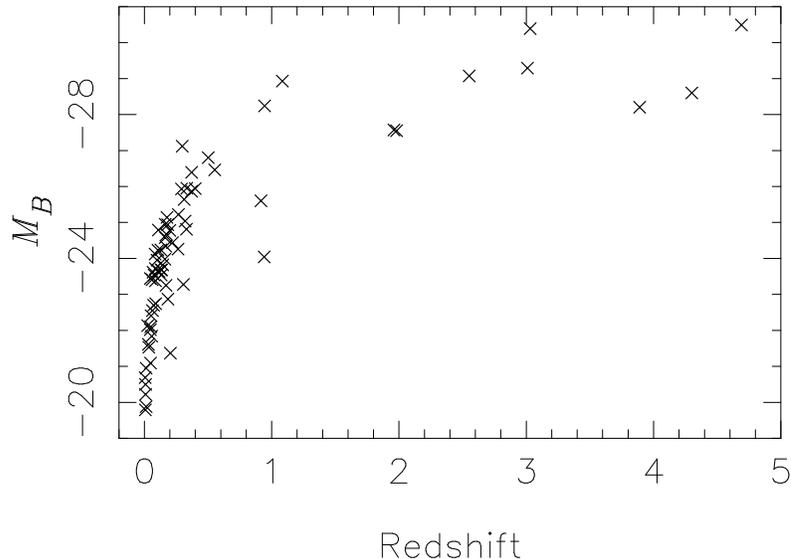}
\caption{Absolute blue magnitude (H$_0 = 50$ km
 s$^{-1}$ Mpc$^{-1}$ and q$_0 = 0.5$)
against redshift for the 72 quasars
in the sample.}
\label{fig:zMB}
\end{figure}

The final sample for the US Key Project consists of 72 quasars
observed with the ISOPHOT instrument (Lemke et al. 1996) in most or
all of the following bands: 5, 7, 12, 25, 60, 100, 135, and 200
$\mu$m.  Ninety percent of the quasars in the sample have redshifts $z
< 1$, while the remaining 10\% lie in the range $2 < z < 4.7$ (Figure
\ref{fig:zMB}).  More than half of the sample consists of luminous
X-ray sources, 25\% are strong UV emitters, and smaller subgroups
contain strong infrared sources, X-ray-quiet objects, red quasars, and
BALQSOs.  Data collection was completed with the end of the
longer-than-expected mission in April, 1998, and analysis is ongoing.
The infrared data points will be combined with all available fluxes at
other wavebands to generate a comprehensive atlas of broadband SEDs
(see Wilkes 1997 for some examples).

The ISO mission, particularly the ISOPHOT instrument, has suffered
from some unanticipated difficulties.  The most egregious of these is
that the instrument sensitivity in some bands is lower than preflight
expectations by a large factor.  This was ameliorated somewhat by the
extended mission lifetime.  The originally preferred technique for
observing faint sources, chopping between the source and the sky, was
abandoned due to concerns about uncertainties in interpreting and
calibrating chopped measurements, particularly at long wavelengths.
Our program was switched mid-stream from chopping to small raster
scans.  Chopped data were obtained for 53 targets, 18 were reobserved
in raster mode, plus 19 new sources were observed only with raster
scans.  The change in observing strategy, combined with the lost
sensitivity, has resulted in a halving of the originally planned
sample.  However, we now have the added benefits of data from both
observing modes for a subset of the targets and better information
about background variations from the raster maps.

\section{The Data}

Reduction of faint object data taken with the ISOPHOT instrument has
been complex and somewhat uncertain, due to factors including the
difficulties mentioned above, the effects of cosmic rays, a very large
parameter space which needs to be calibrated, and the familiarization
time required for any new instrument.  However, the ISOPHOT instrument
team, the ISOPHOT Data Center, and the PHOT Interactive Analysis (PIA)
software team have worked diligently to bring the reduction process to
the point where reliable fluxes for some quasars are being produced
(Haas et al. 1998).  While our results at this stage are still
preliminary, current progress in reduction technique and calibrations
is rapid.

\begin{figure}
\psfig{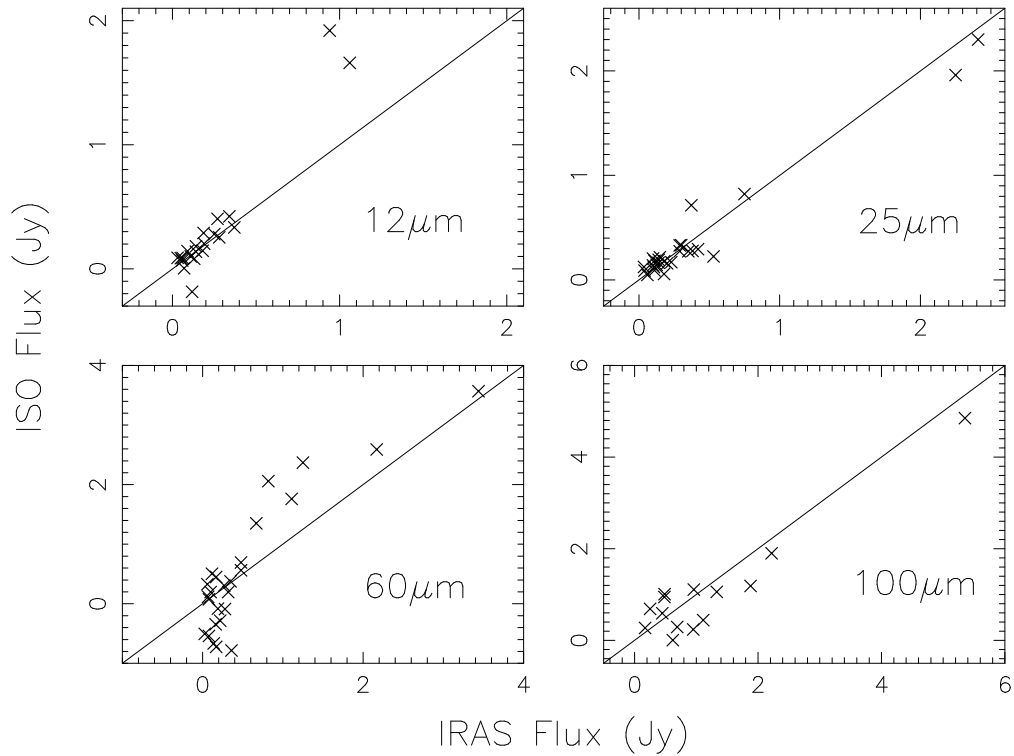}
\caption{Comparison of ISO and IRAS fluxes for chopped measurements.
Lines corresponding to equal flux observed with both telescopes are provided
to guide the eye.}
\label{fig:iso_iras}
\end{figure}

The bulk of the data reduction, including most instrumental
calibrations and corrections, is done with PIA, the standard software
for PHOT reductions (Gabriel, Acosta-Pulido, \& Heinrichsen 1998).
Multiple techniques are being explored for the final step, extracting
the source flux.  Consequently we use scripts written by members of
the ISOPHOT Data Center (IDC) which have not yet been incorporated
into PIA.  Two methods for processing chopped data are currently
utilized: subtracting the average of adjacent sky measurements from
each on-source value in the observing sequence; and Fourier analysis
to look for a significant signal at the chopping frequency.  Small
maps can be generated from the raster scans using PIA, but a simpler
technique of subtracting the flux at the source position from average
adjacent sky values for each pixel has proven to be easier to
interpret and possibly more sensitive.

Figure \ref{fig:iso_iras} shows a comparison of ISO fluxes from
chopped observations reduced in PIA batch mode with the corresponding
IRAS values.  There is general agreement between the two sets of
measurements, particularly at the shorter wavelengths.  However, the
60$\mu$m plot has some disturbing features, notably a large dispersion
about zero for the ISO values and a systematic discrepancy with the
IRAS results at higher flux levels.  The C100 array, which was used
for the 60$\mu$m and 100$\mu$m observations, is known to have had
abnormally high dark current in some pixels.  Cosmic ray strikes can
have residual effects on the data stream which are not fully corrected
or flagged with standard batch reduction.  These and other
instrumental effects are probably responsible for the behavior at
60$\mu$m.  The highest quality final results will require detailed
individual data reduction beyond the batch processing in some cases.

Several issues are being investigated in the pursuit of an optimal
data reduction strategy and set of calibration information.
Measurements of the on-board flux calibration sensors were
occasionally affected by glitches and detector responsivity drift.
Alternative default calibrations are being further refined at the IDC.
In collaboration with workers at the IDC and elsewhere, we are testing
new sets of vignetting corrections, which can have a substantial
effect on the detection of faint sources.  Different techniques for
estimating source flux, taking into account the long-term drift over
the whole observation, are being explored.  More accurate uncertainty
estimates, utilizing all of the available data, will soon be
incorporated into the scripts.

\acknowledgments Martin Haas, Sven M\"{u}ller, Mari Poletta, Ann
Wehrle provided invaluable help with the data reduction.  We
benefitted greatly from discussions with Thierry Courvoisier,
P\'{e}ter \'{A}brah\'{a}m, Ilse van Bemmel, Rolf Chini, and Bill
Reach.  The IDC, IPAC, and the INTEGRAL Science Data Center were very
hospitable during visits to work on this project.  The financial
support of NASA grant NAGW-3134 is gratefully acknowledged.

\end{document}